\documentclass[superscriptaddress,
twocolumn,showpacs,preprintnumbers,amsmath,amssymb]{revtex4}
\usepackage[dvips]{graphicx}
\usepackage{epsf}
\usepackage{epsfig}
\usepackage{latexsym}
\usepackage{amssymb}
\usepackage{amsfonts,amsbsy}
\usepackage{amsmath}
\usepackage{dcolumn}
\usepackage{bm}
\usepackage{pifont}
\begin{document}
\title{Intra-cellular transport of single-headed molecular motors KIF1A}
\author{Katsuhiro Nishinari}
\affiliation{%
Department of Aeronautics and Astronautics,
Faculty of Engineering, University of Tokyo,
Hongo, Bunkyo-ku, Tokyo 113-8656, Japan.
}%
\author{Yasushi Okada}%
 \affiliation{%
Department of Cell Biology and Anatomy, Graduate School of Medicine
University of Tokyo, Hongo, Bunkyo-ku,
Tokyo 113-0033, Japan.
}%
\author{Andreas Schadschneider}%
\affiliation{%
Institut  f\"ur Theoretische  Physik, Universit\"at 
zu K\"oln D-50937 K\"oln, Germany
}%
\author{Debashish Chowdhury}
\affiliation{%
Department of Physics, Indian Institute of Technology,
Kanpur 208016, India.
}%
\date{\today}
\begin{abstract}
Motivated by experiments on single-headed kinesin KIF1A, we develop
a model of intra-cellular transport by interacting molecular motors.
It captures explicitly not only the effects of ATP hydrolysis, but
also the ratchet mechanism which drives individual motors. Our model 
accounts for the experimentally observed single molecule properties 
in the low density limit and also predicts a phase diagram that shows 
the influence of hydrolysis and Langmuir kinetics on the collective 
spatio-temporal organization of the motors. Finally, we provide 
experimental evidence for the existence of domain walls in our 
{\it in-vitro} experiment with fluorescently labeled KIF1A.
\end{abstract}
\pacs{87.16.Nn, 
45.70.Vn, 
02.50.Ey, 
05.40.-a 
}
\maketitle

Intra-cellular transport of a wide variety of cargo in eucaryotic cells 
is made possible by motor proteins, like kinesin and dynein, which move 
on filamentary tracks called microtubules (MT) \cite{howard,schliwa}. 
However, often a single MT is used simultaneously by many motors and, in 
such circumstances, the inter-motor interactions cannot be ignored. 
Fundamental understanding of these collective physical phenomena may also 
expose the causes of motor-related diseases (e.g., Alzheimer's disease) 
\cite{hirotaked} thereby helping, possibly, also in their control and 
cure. Some of the most recent theoretical models of interacting molecular 
motors \cite{frey,santen,popkov,lipo} utilize the similarities between 
molecular motor traffic on MT and vehicular traffic on highways \cite{css} 
both of which can be modelled by appropriate extensions of driven 
diffusive lattice gases \cite{sz,schuetz}. In those models the motor is 
represented by a self-driven particle and the dynamics of the model is 
essentially an extension of that of the asymmetric simple exclusion 
processes (ASEP) \cite{sz,schuetz} that includes Langmuir-like kinetics 
of adsorption and desorption of the motors. In reality, a motor protein 
is an enzyme whose mechanical movement is loosely coupled with its 
biochemical cycle. In this letter we consider specifically the 
{\it single-headed} kinesin motor, KIF1A \cite{okada1,okada3,unpub,Nitta}; 
the movement of a single KIF1A motor has been modelled recently with a 
Brownian ratchet mechanism \cite{julicher,reimann}. In contrast to the 
earlier models \cite{frey,santen,popkov,lipo} of molecular motor traffic, 
which take into account only the mutual interactions of the motors, our 
model explicitly incorporates also the Brownian ratchet mechanism of 
individual KIF1A motors, including its biochemical cycle that involves 
{\it adenosine triphosphate(ATP) hydrolysis}.

The ASEP-like models successfully explain the occurrence of shocks. 
But since most of the bio-chemistry is captured in these models through 
a single effective hopping rate, it is difficult to make direct 
quantitative comparison with experimental data which depend on such 
chemical processes. In contrast, the model we propose 
incorporates the essential steps in the biochemical processes of KIF1A 
as well as their mutual interactions and involves parameters that have 
one-to-one correspondence with experimentally controllable quantities. 

\begin{figure}[tb]
\begin{center}
\includegraphics[width=0.4\textwidth]{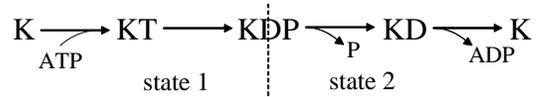}\\
\end{center}
\caption{The biochemical and mechanical states of a single KIF1A motor.
On the left of the dotted line, KIF1A is bound to a fixed position on 
the MT (state 1), while on the right it diffuses along the MT track 
(state 2). At the transition from state 1 to 2, KIF1A detaches from the MT.}
\label{fig-cycle}
\end{figure}

The biochemical processes of kinesin-type molecular motors can be
described by the four states model shown in Fig.~\ref{fig-cycle} 
\cite{okada1,Nitta}: bare kinesin (K), kinesin bound with ATP (KT), 
kinesin bound with the products of hydrolysis, i.e., adenosine 
diphosphate(ADP) and phosphate (KDP), and, finally, kinesin bound with 
ADP (KD) after releasing phosphate. Recent experiments \cite{okada1,Nitta} 
revealed that both K and KT bind to the MT in a stereotypic manner 
(historically called ``strongly bound state'', and here we refer to 
this mechanical state as ``state 1''). KDP has a very short lifetime and 
the release of phosphate transiently detaches kinesin from MT \cite{Nitta}.
Then, KD re-binds to the MT and executes Brownian motion along the 
track (historically called ``weakly bound state'', and here referred to 
as ``state 2''). Finally, KD releases ADP when it steps forward to the 
next binding site on the MT utilizing a Brownian ratchet mechanism, and 
thereby returns to the state K.

\begin{figure}[tb]
\begin{center}
\includegraphics[width=0.45\textwidth]{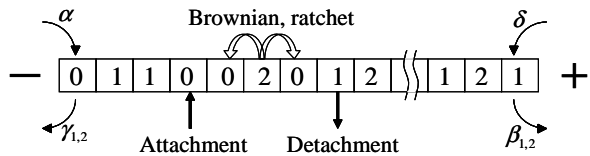}\\
\end{center}
\caption{A 3-state model for molecular motors moving along a MT. 
0 denotes an empty site, 1 is K or KT and 2 is KD. Transition from 1 to 2, 
corresponding to hydrolysis, occurs within a site whereas movement to 
the forward or backward site occurs only when motor is in state 2. At 
the minus and plus ends the probabilities are different from those in 
the bulk.}
\label{fig2}
\end{figure}

{\it Model definition.} ---
A single protofilament of MT is modelled by a one-dimensional lattice of 
$L$ sites each of which corresponds to one KIF1A-binding site on the MT; 
the lattice spacing is equivalent to $8$ nm which is the separation 
between the successive binding sites on a MT \cite{howard}. Each kinesin 
is represented by a  particle with two possible internal states labelled 
by the indices $1$ and $2$. Attachment of a motor to the MT occurs 
stochastically whenever a binding site on the latter is empty. 
Attachment and detachment at the two ends of the lattice need 
careful treatment and will be specified below. Thus, each of the lattice 
sites can be in one of three possible allowed states (Fig.~\ref{fig2}): 
empty (denoted by $0$), occupied by a kinesin in state $1$, or occupied 
by a kinesin in state $2$. 

For the dynamical evolution of the system, one of the $L$ sites is 
picked up randomly and updated according to the rules given below 
together with the corresponding probabilities (Fig.~\ref{fig2}):
\begin{eqnarray}
 &&{\rm Attachment:} \,\,\,\,\,  0\to 1 \,\,\,{\rm with} \,\, \omega_a dt\\
 &&{\rm Detachment:} \,\,\,\, 1\to 0 \,\,\, {\rm with} \,\, \ \omega_d dt\\
 &&{\rm Hydrolysis:} \,\,\,\,\,  1\to 2 \,\,\,{\rm with} \,\, \omega_h dt\\
 &&{\rm Ratchet:}\,\,\,\,\, \left\{\begin{array}{c}
     2 \to 1\,\,\,{\rm with} \,\, \omega_s dt\\
     20 \to 01\,\,\,{\rm with} \,\, \omega_f dt
   \end{array}\right.\\
 &&{\rm Brownian\ motion:}\,\,\,\,\, \left\{\begin{array}{c}
     20 \to 02\,\,\,{\rm with} \,\, \omega_b dt\\
     02 \to 20\,\,\,{\rm with} \,\, \omega_b dt
   \end{array}\right.
\end{eqnarray}

The probabilities of detachment and attachment at the two ends of 
the MT may be different from those at any bulk site. 
We choose $\alpha$ and $\delta$, instead of $\omega_a$, as the 
probabilities of attachment at the left and right ends, respectively. 
Similarly, we take $\gamma_1$ and $\beta_1$, instead of $\omega_d$, 
as probabilities of detachments at the two ends (Fig.~\ref{fig2}). Finally, 
$\gamma_2$ and $\beta_2$, instead of $\omega_b$, are the probabilities 
of exit of the motors through the two ends by random Brownian movements.

Let us relate the rate constants $\omega_f$, $\omega_s$ and $\omega_b$ 
with the corresponding physical processes in the Brownian ratchet 
mechanism of a single KIF1A motor. Suppose, just like models of 
flashing ratchets \cite{julicher,reimann}, the motor ``sees'' a
time-dependent effective potential which, over each biochemical cycle, 
switches back and forth between (i) a periodic but asymmetric sawtooth 
like form and (ii) a constant. The rate constant $\omega_h$ in our 
model corresponds to the rate of the transition of the potential from 
the form (i) to the form (ii). The transition from (i) to (ii) happens 
soon after ATP hydrolysis, while the transition from (ii) to (i) happens 
when ATP attaches to a bare kinesin\cite{okada1}. The rate constant 
$\omega_b$ of the motor in state $2$ captures the Brownian motion of the 
free particle subjected to the flat potential (ii). The rate constants 
$\omega_s$ and $\omega_f$ are proportional to the overlaps of the 
Gaussian probability distribution of the free Brownian particle with, 
respectively, the original well and the well immediately in front of the 
original well of the sawtooth potential. 

Let us denote the probabilities of finding a KIF1A molecule in 
the states $1$ and $2$ at the lattice site $i$ at time $t$ by the 
symbols $r_i$ and $h_i$, respectively. In mean-field approximation the 
master equations for the dynamics of motors in the bulk of the system 
are given by
\begin{eqnarray}
\frac{dr_i}{dt}&=&\omega_a (1-r_i-h_i) -\omega_h r_i -\omega_d r_i
+\omega_s h_i\nonumber\\
&&+\omega_f h_{i-1}(1-r_i-h_i),\\
\frac{dh_i}{dt}&=&-\omega_s h_i +\omega_h r_i 
-\omega_f h_i (1-r_{i+1}-h_{i+1}) \nonumber\\
&&-\omega_b h_i (2-r_{i+1}-h_{i+1}-r_{i-1}-h_{i-1})  \nonumber\\
&&+\omega_b (h_{i-1}+h_{i+1})(1-r_i-h_i). 
\label{eq-bulk}
\end{eqnarray}
The corresponding equations for the boundaries, which depend on the rate 
constants $\alpha$, $\delta$, $\gamma_i$ and $\beta_i$ for entry and 
exit (Fig.~\ref{fig2}), are similar and will be presented elsewhere 
\cite{unpub}.

From experimental data \cite{okada1,okada3}, good estimates for the 
parameters of the suggested model can be obtained. Assuming that one 
timestep corresponds to 1~ms, each simulation run had a duration
of 1 minute in real time.  The length of MT is fixed 
as $L=600$. The detachment rate $\omega_d \simeq 0.0001$ ms$^{-1}$ is 
found to be independent of the kinesin population. On the other hand, 
$\omega_a = 10^7$~$C$/M$\cdot$s depends on the concentration $C$ (in M) 
of the kinesin motors. In typical eucaryotic cells {\it in-vivo} the 
kinesin concentration can vary between 10 and 1000 nM. Therefore, the 
allowed range of $\omega_a$ is 
$0.0001$ ms$^{-1} \leq \omega_a \leq 0.01$ ms$^{-1}$. 
The rate $\omega_b^{-1}$ must be such that the Brownian diffusion 
coefficient $D$ in state 2 is of the order of $40000$~nm$^2$/s; 
using the the relation 
$\omega_b \sim D/(8\text{nm})^2$, we get $\omega_b \simeq 0.6$~ms$^{-1}$. 
Moreover, from the experimental observations that 
$\omega_f/\omega_s \simeq 3/8$ and 
$\omega_s + \omega_f \simeq 0.2$~ms$^{-1}$, 
we get the individual estimates
$\omega_s \simeq 0.145$~ms$^{-1}$ and $\omega_f \simeq 0.055$~ms$^{-1}$. 
The experimental data on the Michaelis-Menten type kinetics of
hydrolysis \cite{howard} suggest that
\begin{equation}
\omega_h^{-1} \simeq \biggl[ 4 + 9 \biggl(
\frac{0.1~\text{mM}}{\text{ATP concentration (in 
mM)}}\biggr)
\biggr] \text{ms}
\label{MM}
\end{equation}
so that the allowed biologically relevant range of 
$\omega_h$ is $0 \leq \omega_h \leq 0.25$~ms$^{-1}$.


{\it Single-molecule properties.} ---
An important test for the model is provided by a quantitative
comparision of the low density properties with empirical results.
Single molecule experiments \cite{okada1} on KIF1A have established that \\
(i) $v$, the mean speed of the kinesins, is about $0.2$~nm/ms if 
the supply of ATP is sufficient, and that $v$ decreases with the lowering 
of ATP concentration following a Michaelis-Menten type relation like (\ref{MM}); \\
(ii) $D/v \sim 190$~nm, irrespective of the ATP concentration, 
where $D$ is the diffusion constant; \\
(iii) $\tau$, the mean duration of the movement of a kinesin on the 
MT, is more than $5$~s, irrespective of the ATP concentration.\\
The corresponding predictions of our model (see Table~\ref{tab-1mol}) for 
$\omega_{a} = \alpha = 1.0 \times 10^{-6}$ ms$^{-1}$, which allows 
realization of the condition of low density of kinesins, are in excellent 
agreement with the experimental results.

\begin{table} 
\begin{tabular}{|c|c|c|c|c|} \hline
ATP (mM)&  $\omega_h$ (1/ms)& $v$ (nm/ms)&  $D/v$ (nm) & $\tau$ (s)\\\hline
$\infty$ & 0.25 & 0.201 & 184.8 & 7.22 \\ \hline
0.9      & 0.20 & 0.176 & 179.1 & 6.94 \\ \hline
0.3375   & 0.15 & 0.153 & 188.2 & 6.98 \\ \hline
0.15     & 0.10 & 0.124 & 178.7 & 6.62 \\ \hline
\end{tabular}
\caption{\label{tab-1mol}{Predicted transport properties 
from this model in the low-density limit for four different ATP densities.
$\tau$ is calculated by averaging the intervals between attachment
 and detachment of each KIF1A.}}
\end{table} 

{\it Collective properties.} ---
Assuming {\em periodic} boundary conditions, the solutions 
$(r_i, h_i)=(r,h)$ of the mean-field equations
(\ref{eq-bulk}) in the steady-state 
are found to be 
\begin{eqnarray} 
r&=&\frac{ -\Omega_h - \Omega_s -  (\Omega_s -1)K  + 
{\sqrt{D}} }{2 K(1+K)},\label{eqr}\\
h&=&\frac{\Omega_h +\Omega_s + (\Omega_s +1)K  
-{\sqrt{D}} }{2 K}\label{eqh}
\end{eqnarray}
where $K=\omega_d/\omega_a$,
$\Omega_h=\omega_h/\omega_f$, $\Omega_s=\omega_s/\omega_f$, and
\begin{equation}
 D=4\Omega_s K(1+K)+
{\left( \Omega_h +\Omega_s + ( \Omega_s-1)K  \right) }^2.
\end{equation}
The probability of finding an empty binding site on a MT is $Kr$ as 
the stationary solution satisfies the equation $r+h+Kr=1$. The 
steady-state flux of the motors along their MT tracks is then given 
by $J=\omega_f h(1-r-h).$ It is interesting to note that in the 
{\it low} ATP concentration limit 
($\omega_h \ll \omega_s \simeq \omega_f$) of our model, the flux of 
the motors is well approximated by 
$J_{\rm low} = q_{\text{eff}} \rho (1-\rho)$, 
which formally looks like the corresponding expression for the totally 
asymmetric exclusion process, where $\rho$ is close to the Langmuir 
limit $1/(1+K)$ and, 
\begin{equation}
q_{\text{eff}} = \frac{\omega_h (1+K)}{\Omega_s (1+K) + K} 
\end{equation}
as the effective hopping probability\cite{unpub}. 

\begin{figure}[htb]
\begin{center}
\includegraphics[width=0.45\columnwidth]{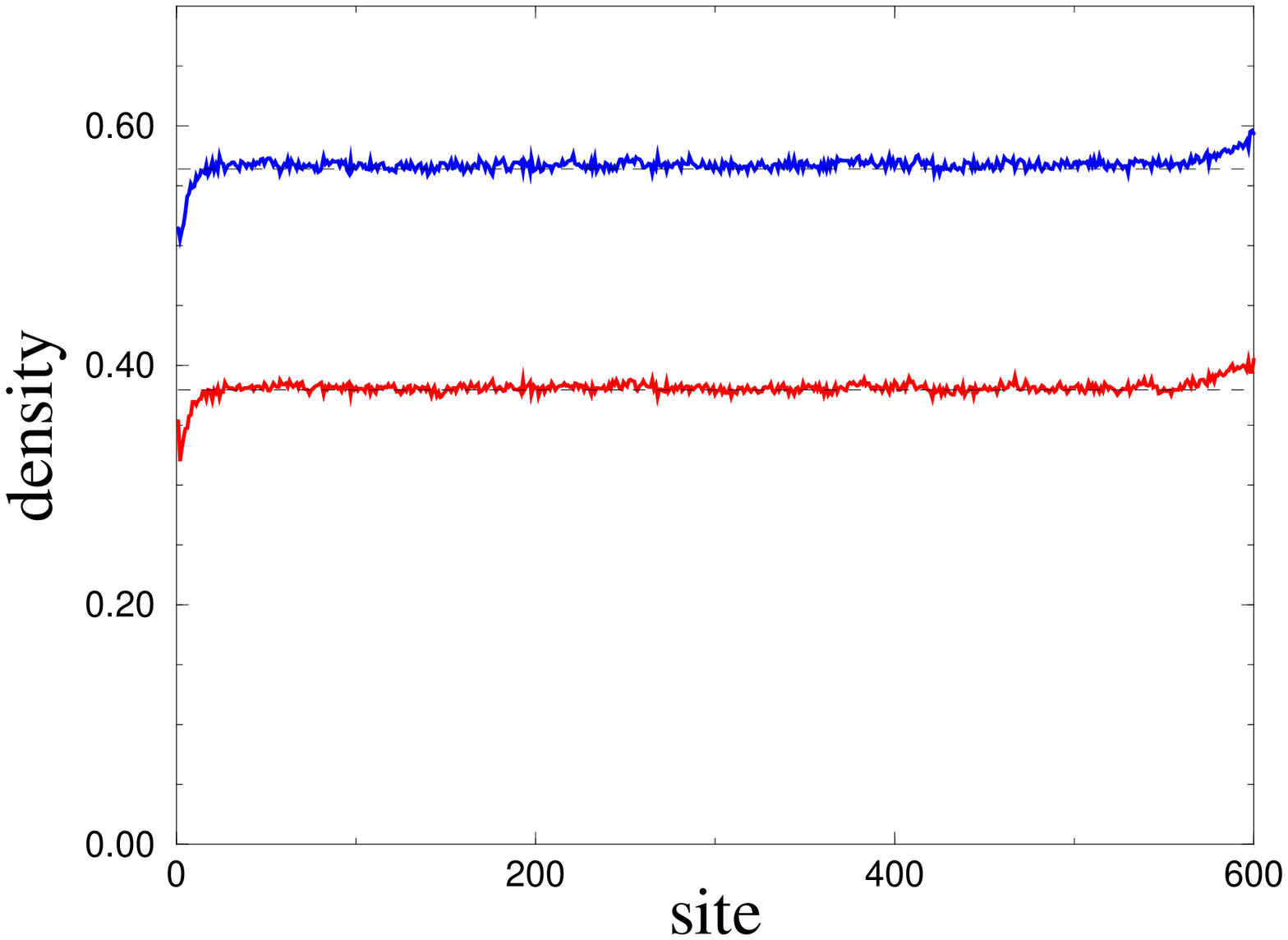}
\includegraphics[width=0.45\columnwidth]{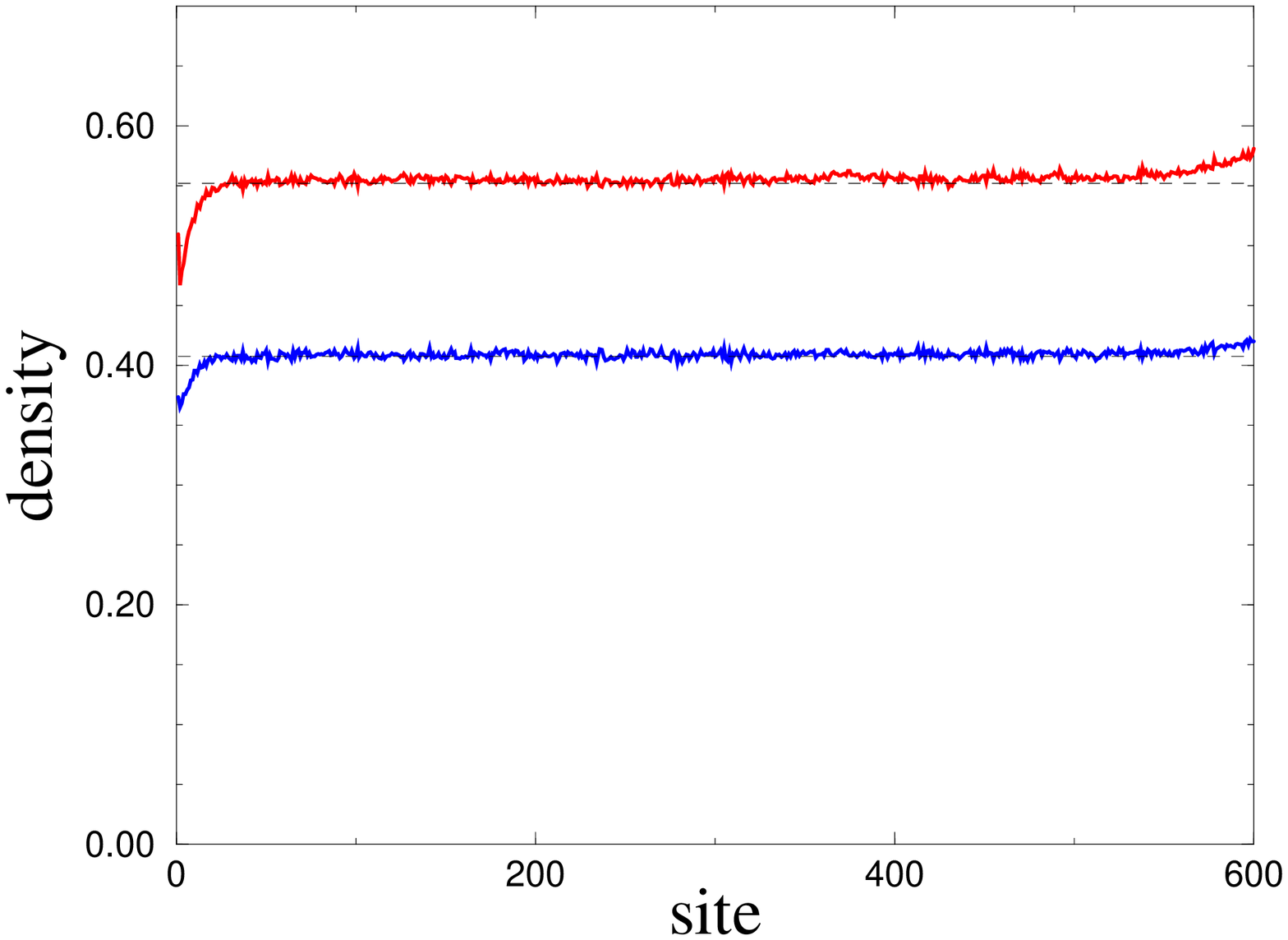}
\end{center}
\caption{The stationary density profiles for $\omega_h=0.1$ (left) and 
$\omega_h=0.2$ (right) in the case $\omega_a=0.001$.
The blue and red lines correspond to the densities of state 1 and 2,
respectively. The dashed lines are the mean-field predictions (\ref{eqr})
and (\ref{eqh}) for periodic systems with the same parameters.
}
\label{fig-density}
\end{figure}

\begin{figure}[htb]
\begin{center}
\includegraphics[width=0.9\columnwidth]{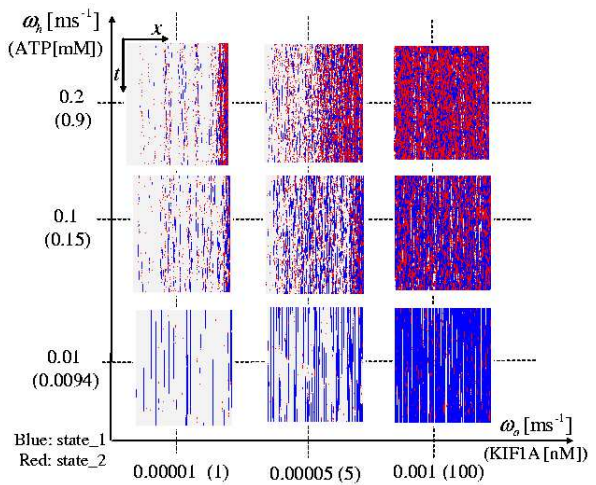}
\end{center}
\caption{Phase diagram of the model in the $\omega_h-\omega_a$ plane,
with the corresponding values for ATP and KIF1A concentrations
given in brackets.  These quantities are controllable in experiment.
The boundary rates are $\alpha=\omega_a,
 \beta_{1,2}=\omega_d, \gamma_{1,2}=\delta=0$.
The position of the immobile shock depends on both ATP and KIF1A 
concentrations.
}
\label{fig-phasediag}
\end{figure}

Although the system with periodic boundary conditions is fictitious, 
the results provide good estimates of the density and flux in the 
corresponding system with open boundary conditions, particularly, in 
the high $\omega_a$ regime (Fig.~\ref{fig-density}) which corresponds 
to jammed traffic of kinesin on MT (see Fig.~\ref{fig-phasediag}). 
We also see that, for a given $\omega_a$, the bulk density of motors 
in state 2 exceeds that of those in state 1 as $\omega_h$ increases 
beyond a certain value.

{\it Phase diagram.} ---
In contrast to the phase diagrams in the $\alpha-\beta$-plane reported 
by earlier investigators \cite{frey,santen,lipo}, we have drawn the 
phase diagram of our model (Fig.~\ref{fig-phasediag}) in the 
$\omega_a-\omega_h$ plane by carrying out extensive computer simulations 
for realistic parameter values of the model with open boundary conditions. 
The phase diagram shows the strong influence of hydrolysis on
the spatial distribution of the motors along the MT. For very low 
$\omega_h$ no kinesins can exist in state 2; the kinesins, all of which 
are in state 1, are distributed rather homogeneously over the entire 
system. In this case the only dynamics present is due to the Langmuir 
kinetics.

Even a small, but finite, rate $\omega_h$ is sufficient to change this 
scenario. In this case both the density profiles $\rho_j^{1}$ and 
$\rho_j^{2}$ of kinesins in the states 1 and 2 exhibit a shock. 
As in the case of the ASEP-like models with Langmuir kinetics 
\cite{frey,santen}, these shocks are localized. 
In computer simulations we have observed that the shocks in density 
profiles of kinesins in the states 1 and 2 always appear at the {\em same}
position. 
Note that if the individual density profiles $\rho_j^{1}$ and 
$\rho_j^{2}$ exhibited shocks at two different locations, two shocks 
would appear in the {\it total} density profile 
$\rho_j=\rho_j^{1}+ \rho_j^{2}$ violating the usual arguments 
\cite{shockform} that ASEP-type models exhibit exactly one shock.
Moreover, we have found that the position of the immobile shock 
depends on the concentration of the motors as well as that of ATP; 
the shock moves towards the minus end of the MT with the increase of 
the concentration of kinesin or ATP or both (Fig.~\ref{fig-phasediag}). 

\begin{figure}[t]
\begin{center}
\vspace{0.5cm}
\includegraphics[width=0.7\columnwidth]{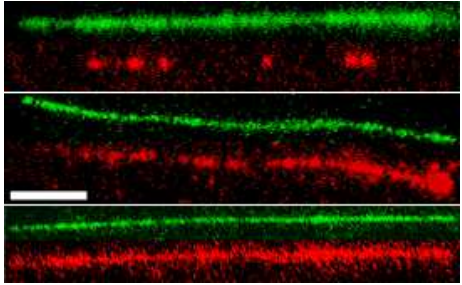}
\end{center}
\caption{Formation of comet-like accumulation of kinesin at the end of
 MT.  Fluorescently labeled KIF1A (red) was introduced to MT (green) 
at 10 pM (top), 100 pM (middle) and 1000 pM (bottom) concentrations 
along with 2 mM ATP.  The length of the white bar is 2$\mu m$.
}
\label{fig-comet}
\end{figure}
Finally, we present direct experimental evidence that support of the 
formation of the shock. The ``comet-like structure'', shown in the 
middle of Fig.~\ref{fig-comet}, is the collective pattern formed by 
the red fluorescent labelled kinesins where a domain wall separates 
the low-density region from the high-density region. The position of 
the domain wall depends on both ATP and KIF1A concentrations. Moreover, 
as we increase the concentration of KIF1A, the transition from the 
regime of free flow of kinesins to the formation of the shock is 
observed(top and middle in Fig.~\ref{fig-comet}). Furthermore, we 
observe jammed traffic of kinesins at sufficiently high concentration 
(bottom in Fig.~\ref{fig-comet}). The position of the shock in our 
simulation agrees well with the location of the domain wall in the 
comet-like structure observed in experiments\cite{unpub}.

In this letter we have developed a stochastic model for the collective 
intra-cellular transport by KIF1A motors, by taking into account the 
biochemical cycle of individual motors involving ATP hydrolysis and their 
mutual steric interactions. We have been able to identify the biologically 
relevant ranges of values of all the model parameters from the empirical 
data. In contrast to some earlier oversimplified models, the predictions 
of our model are in good quantitative agreement with the corresponding 
experimental data. Moreover, we have mapped the phase diagram of the 
model in a plane spanned by the concentrations of ATP and KIF1A, both of 
which are experimentally controllable quantities. Finally, we have reported 
the experimental observation of a comet-like collective pattern formed by 
the kinesin motors KIF1A and identified the domain wall in the pattern 
with the shock predicted by our model. 


\end{document}